\documentclass[proceedings]{rmaa}

\title{X-ray plasma diagnostics for totally and partially photoionized plasmas such as Warm Absorber in AGN}

\author{Delphine Porquet\altaffilmark{1}
  and Jacques Dubau\altaffilmark{2}
  \affil{Observatoire de Paris, section de Meudon (DAEC$^{1}$, DARC$^{2}$), France}
        } 

\fulladdresses{
\item Delphine Porquet \& Jacques Dubau: Observatoire de Paris,
  section de Meudon,
  92195 Meudon Cedex,
  France (delphine.porquet@obspm.fr, Jacques.Dubau@obspm.fr)
  }
\shortauthor{Porquet \& Dubau}
\shorttitle{X-ray plasmas diagnostics}

\SetVolume{99} \SetFirstPage{167} \SetYear{2000}
\ReceivedDate{1999 December 1} 
\AcceptedDate{1999 December 1} 

\abstract{Thanks to the new generation of X-ray satellites such as Chandra and XMM, high resolution and high sensitivity spectra are available. In particular, for the first time, the three most intense lines (resonance, intercombination and forbidden) of low charged (low Z) He-like ions are splitted for non solar plasmas.\\
We present density, ionizing process and temperature diagnostics, for totally and partially photoionized plasmas, based on ratios of these three lines. These powerful plasma diagnostics could be used for hot astrophysical plasmas such as AGN, starburst galaxies, X-ray binaries, etc. In particular, they could be applied to Warm Absorber often seen in Active Galactic Nuclei (Porquet \& Dubau \cite{PorquetDubau2000}), which is  an important key tool to understand central region of different types of AGN (Seyfert\,1 and 2, high and low redshift quasars).}      

\begin{document}

\maketitle

\section{A brief introduction to Warm Absorber}\label{sec:WA}
Warm Absorber (WA) has been found, few years ago, to be an important component of the central region of Active Galactic Nuclei. Indeed, it is supposed to be located between the Broad Line Region (BLR) and the Narrow Line Region (NLR), and even inside the BLR (Otani et al. \cite{Otani96}, Porquet et al. \cite{Porquet99}). This medium is supposed to be a photoionized plasma, but some additional ionization process is not ruled out (Porquet \& Dumont \cite{PorquetDumont98}, Porquet et al. \cite{Porquet99}). It is why, we can consider that Warm Absorber could be either totally or partially photoionized (photoionization plus collisional ionization). The existence of WA has been revealed by observations of significant absorption edges, in the X-ray spectra near 0.8\,keV, which imply column densities of 10$^{21}$--10$^{23}$\,cm$^{-2}$. This features are seen in at least 50$\%$ of Seyfert\,1 (Reynolds \cite{Reynolds97}), thus WA is a common characteristics of these objects. This medium is supposed to be not only an absorber but moreover a multi-wavelength emitter (such as the optical coronal lines: Porquet et al. \cite{Porquet99}). Besides, in Seyfert galaxies soft X-ray emission lines are also observed. In particular, the He-like X-ray lines are of particular interest since their ratios give plasma diagnostics. This is the topic of the next section.\\
As already said, WA being either totally or partially photoionized, we have thus to study the two following ionization models: totally photoionized plasmas (``pure'' photoionization) and hybrid plasmas (photoionization plus an additional ionization process).

\section{X-ray plasma diagnostics}\label{sec:diagnostics}

He-like ions have an interesting atomic structure, they emit three main lines (n=2 shell)  which are close in wavelengths: resonance (called w), intercombination (x+y) and forbidden lines (z). As shown by Gabriel \& Jordan \cite{Gabriel69}, the combination of the ratio of these lines can be used for electronic density (n$_{\mathrm{e}}$) and electronic temperature (T$_{\mathrm{e}}$) diagnostics:
\begin{equation}\label{eq:R}
\mathrm{R(n_{e})=\frac{z}{x+y}~~~and~~~~G(T_{e})=\frac{(x+y)+z}{w}}
\end{equation}
As pointed out by Pradhan \cite{Pradhan85} and Liedahl \cite{Liedahl99}, these diagnostics could also be used for the study of photoionized plasmas.  

\subsection{Atomic data}\label{sec:atomicdata}

In order to obtain accurate line ratios, we have calculated atomic data over a wide range of temperature for radiative and dielectronic recombinations\footnote{Radiative recombination dominates at low temperature whereas dielectronic recombination dominates at high temperature (considering recombination processes).}, and collisional excitation\footnote{Collisional excitations inside the n=2 shell occur even at low temperature.} from the ground level, which should be considered in hybrid plasmas (high temperature plasmas).\\
For both atomic processes (recombination and excitation), radiative cascades from n$>$2 levels have been taken into account in the calculation of the population of the n=2 shell levels (related to w, x+y, z). Indeed, cascades are of great importance especially for the z forbidden lines. The rate coefficients and more details about the calculations can be found in Porquet \& Dubau \cite{PorquetDubau2000}.

\subsection{Ionizing process diagnostic}\label{sec:ionizationprocess}

In ``purely'' photoionized plasmas, the radiative recombination is the only ionization process or is the most dominant. In the so-called hybrid plasma, the collisional excitation is no more negligible since temperature is high enough to permit excitations from the ground level.\\
Radiative recombination rates show that the level connected to the intercombination (x+y) and to the forbidden (z) lines are favored and thus implies stronger triplet lines (x,y,z) than the singlet resonance line (w). On the contrary, collisional strengths favored the level connected to the w resonance line.\\
Then, totally photoionized media emit weak resonance (w) line compared to the forbidden (z) or to the intercombination (x+y) lines. In hybrid plasmas, the behaviour is the opposite (see fig~\ref{fig1}).

\subsection{Density diagnostic}\label{sec:densitydiag}

\begin{figure}[h]
\begin{center}
\includegraphics[width=7cm]{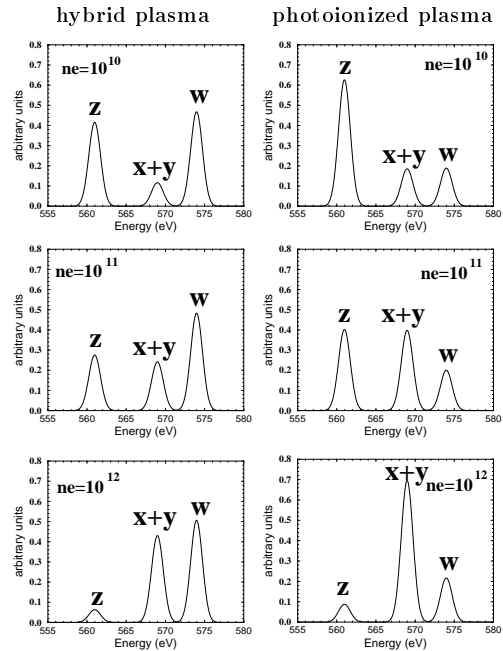}
\end{center}
\caption{O\,VII theoretical spectra constructed using the RGS (XMM) resolving power (E/$\Delta$E) for three density values (in cm$^{-3}$). This corresponds (approximatively) to the range where the ratio R is very sensitive to the density. z: forbidden lines, x+y: intercombination lines and w: resonance line.}
\label{fig1}
\end{figure}

The ratio R is constant below some critical density (n$_{\mathrm{crit}}$), the forbidden line is then intense, and above n$_{\mathrm{crit}}$, the $^{3}$S$_{1}$ level (forbidden line) is depopulated to the $^{3}$P levels (intercombination lines) via collisional excitation inside n=2. Then, the intensity of the forbidden (z) line decreases while the intensity of the intercombination (x+y) lines increases. This inversion occurs inside (approximatively) two density magnitudes. Thus inside this range, R is very sensitive to the density and gives accurate estimation of the density (see fig~\ref{fig1}).\\
Thus to summarize, when we observe a ratio R equals to the region where R is constant, we obtain an upper limit for the density. Inside the two magnitudes where R becomes sensitive to the density, we obtain the density. And, when R tends to zero (z$\to$0) we obtain a lower limit for the density.

\section{Conclusion}

For the study of Warm Absorber, we propose two powerful diagnostics for electron temperature and density. The first of them could be used to determine its ionizing process: totally or partially photoionized medium. Calculations of atomic data and line ratios are available in Porquet \& Dubau \cite{PorquetDubau2000}. The determination of the physical parameters of WA in different types of AGN will have a great impact on unified models.

\end{document}